\documentclass[onecolumn,11pt]{article}

\newcommand{\bq}{\begin{quote}}
\newcommand{\eq}{\end{quote}}
\newcommand{\ben}{\begin{enumerate}}
\newcommand{\een}{\end{enumerate}}
\title{QBism, Bohr, and the quantum omelette \\
tossed by de Ronde}
\author{Ulrich Mohrhoff\\
\textit{\small Sri Aurobindo International Centre of Education}\\ 
\textit{\small Pondicherry 605002 India}\\
\ttfamily{\small ujm@auromail.net}}
\date{}
\normalsize
\begin{document}
\maketitle
\begin{abstract}
\noindent In his recent paper ``QBism, FAPP and the Quantum Omelette,'' de Ronde makes a variety of questionable claims concerning QBism, Bohr, and the present author's critical appraisal of QBism. These claims are examined. Subsequently an outline is presented of what one might see if one looks into the quantum domain through the window provided by the quantum-mechanical correlations between outcome-indicating events in the classical domain.
\end{abstract}

\vskip2\baselineskip\noindent The formalism of quantum mechanics (QM) was characterized by Jaynes \cite{Jaynes} as ``a peculiar mixture describing in part realities of Nature, in part incomplete human information about Nature---all scrambled up by Heisenberg and Bohr into an omelette that nobody has seen how to unscramble.'' The (alleged) improper scrambling of ontic (``objective'') and epistemic (``subjective'') perspectives is the focus of a recent paper by de Ronde \cite{CdR}. What mainly concerns me here is his defense of QBism, which he regards as ``one of the most honest, consistent and clear approaches to QM'' and as ``completely safe from several (ontological) criticisms it has recently received,'' including one by the present author \cite{Mohrhoff_QBism}, notwithstanding that it ``does not solve the problems of QM, it simply dissolves them.'' In their response to Nauenberg \cite{Nauenberg}, another critic of QBism, the QBist triumvirate Fuchs et al.\ \cite{FMS_Nauenberg} ``welcome criticism, but urge critics to pay some attention to what we are saying.'' I could say the same of de Ronde's throwaway remarks on my critique of QBism.

What follows is divided into two parts. The first part deals with Bohr, QBism, and what de Ronde has to say about Bohr, QBism, and my critical appraisal of QBism. The second part outlines what one might see if one looks into the quantum domain through the window provided by the quantum-mechanical correlations between outcome-indicating events in the classical domain.

\section{}
All de Ronde quotes from my critical appraisal of QBism is the distinction I made between
\begin{itemize}
\item[(I)]a transcendental reality external to the subject, undisclosed in experience, which Kant looked upon as the intrinsically unknowable cause of subjective experience, and
\item[(II)]the product of a mental synthesis---a synthesis based on the spatiotemporal structure of experience, achieved with the help of spatiotemporal concepts, and resulting in an objective reality from which the objectifying subject can abstract itself.
\end{itemize}
It is true that within the Kantian scheme, as de Ronde explains, ``transcendental reality amounts to reality as it is, `the thing in itself'.'' It ought to be noted, however, that I carefully avoided the vacuous expression ``reality as it is'' (in and by itself, out of relation to our experience and our categorial schemes). To acknowledge a transcendental reality is but to recognize that there is more to reality than what is disclosed in human experience and can be captured by mathematical models or mental constructs. An objective reality constructed by us is the one we physicists will ever be concerned with, whether we want it or not, whether or not we think of it (rightly or wrongly but in any case irrelevantly%
\footnote{As Xenophanes observed some twenty-five centuries ago, even if my conceptions represented the world exactly as it is, I could never know that this is the case.}%
) as a faithful representation of  ``reality as it is.''

De Ronde flatly denies ``that QM can be considered in terms of `objective reality' within the Bohrian scheme.'' Why? Because ``[t]he subject cannot abstract himself from the definition of reality provided by QM in terms of waves, particles or even definite valued properties.'' It is news to me that QM provides a definition of (objective) reality, let alone one in terms of possessed definite properties or classical models like waves or particles. Where Bohr is concerned, objective reality%
\footnote{I am of course not speaking of transcendental reality, whose relation to QM did not concern Bohr.}
is made up of two things: (i)~the experimental arrangement---the system preparation, the measurement apparatus, and the indicated outcome---all of which have to be described in ordinary (``classical'') language if we want to be able to communicate ``what we have done and what we have learned'' (\cite[pp.~3, 24]{Bohr-EAPHK} and \cite[pp.~39, 72, 89]{Bohr-APHK}), and (ii)~``statistical laws governing observations obtained under conditions specified in plain language.'' That's all there is to it:  ``the physical content of quantum mechanics is exhausted by its power to formulate'' such laws \cite[p.~12]{Bohr-EAPHK}.

De Ronde argues that because Bohr's notion of complementarity involves a subject's choice, it is inconsistent with an objective conception of reality: ``Physical reality can be only represented in an objective manner if the subject plays no essential role within that representation.'' Was Bohr then mistaken in writing that the ``description of atomic phenomena has \dots\ a perfectly objective character, in the sense that no explicit reference is made to any individual observer,'' and that ``all subjectivity is avoided by proper attention to the circumstances required for the well-defined use of elementary physical concepts'' \cite[pp.~3, 7]{Bohr-EAPHK}? By no means, for it is not the case that complementarity implies a choice.

To illustrate his point, de Ronde considers a double slit experiment, which Subject~1 performs with both slits open and Subject~2 performs with one slit shut. Subject~1 (who, like Subject~2, appears to owe his information about QM to the popular science media) concludes that the ``quantum object'' is a wave, while Subject~2 concludes that it is a particle. For de Ronde this means that quantum reality, giving rise as it does to subject-dependent conclusions about one and the same object, cannot be objective: ``The real (objective) existence of waves and particles cannot be dependent on a (subjective) choice of an experimenter.'' In point of fact, what we are dealing with here is not a single situation involving a subject's choice but two distinct physical situations within a single objective reality. Nor are the objects studied in these experiments either classical waves or classical particles. They are particles only in the sense that they can produce ``clicks'' in counters,%
\footnote{On the inadequacy of \emph{this} language see Ulfbeck and Bohr \cite{UlfBohr} and my \cite{Mohrhoff_clicks}.}
and they are waves only in the sense that the clicks can exhibit interference fringes.

De Ronde's mention of  ``the Bohrian metaphysical premise according to which the description must be given in terms of classical physics by waves or particles'' suggests that he actually believes that this is what Bohr had in mind when he insisted on the use of the language of classical physics. Which in turn suggests that some of the most central tenets of Bohr's philosophy are lost on de Ronde, such as the necessity of defining observables in terms of the experimental arrangements by which they are measured: the ``procedure of measurement has an essential influence on the conditions on which the very definition of the physical quantities in question rests'' \cite{Bohr1935a}. In other words, to paraphrase a famous dictum by Wheeler,%
\footnote{``No elementary phenomenon is a phenomenon until it is a registered (observed) phenomenon'' \cite{Wheeler}.}
no property (of a quantum system) or value (of a quantum observable) is a possessed property or value unless it is a measured property or value. (For extensive discussions of this point see my \cite{Mohrhoff_manifestingQW,Mohrhoff_QMnewlight}.)

``One of the main constituents of the present quantum omelette,'' de Ronde points out, ``is the idea that ‘measurement’ is a process which has a special status within QM.'' While `measurement' has a special status to be sure, it isn't a process. In fact, QM knows nothing about processes. It is about measurement outcomes---actual, possible, or counterfactual ones, performed on the same system at different times or on different systems in spacelike relation---and their correlations. What happens between a system preparation and a measurement is anybody's guess, as the proliferation of interpretations of QM proves. Given a system preparation, QM gives us the probabilities with which outcome-indicating events happen, not processes by which they come about.

Two problems, according to de Ronde, ``make explicit how QM has turned into a `quantum omelette' with no clear limit between an ontological account and an epistemological one''---``two problems in which the intrusion of a choosing subject appears explicitly in the determination of what is considered to be (classically) real---or actual.'' The first is the basis problem, which, so de Ronde, 
\bq
attempts to explain how is Nature capable of making a choice between different incompatible bases. Which is the objective physical process that leads to a particular basis instead of a another one? If one could explain this path through an objective physical process, then the choice of the experimenter could be regarded as well as part of an objective process\dots. Unfortunately, still today the problem remains with no solution within the limits of the orthodox formalism.
\eq
That this problem remains unsolved should not come as a surprise. It is in the nature of pseudo-problems to lack solutions---real solutions, as against gratuitous ones. The reason this problem is a pseudo-problem is that what happens between a system preparation and a measurement is a phenomenon%
\footnote{``[A]ll unambiguous interpretation of the quantum mechanical formalism involves the fixation of the external conditions, defining the initial state of the atomic system concerned and the character of the possible predictions as regards subsequent observable properties of that system. Any measurement in quantum theory can in fact only refer either to a fixation of the initial state or to the test of such predictions, and it is first the combination of measurements of both kinds which constitutes a well-defined phenomenon.'' \cite{Bohr_1939}.}
that cannot be dissected into the unitary evolution of a quantum state and its subsequent ``collapse.'' There is no Nature making choices, whether between bases or between possible outcomes. There is no objective physical process selecting a particular basis. What determines a particular basis is the measurement apparatus. The fact that the apparatus is usually chosen by an experimenter, however, is of no consequence as far as the interpretation of the formalism is concerned. What matters is that the apparatus is needed not only to indicate the possession of a property by a quantum system but also---and in the first place---to make a set of properties available for attribution to the system. Whether it is anyone's intention to obtain a particular kind of information, or whether anyone is around to take cognizance of it, is perfectly irrelevant. 

The second problem, in the words of de Ronde, is this:
\bq
Given a specific basis (context or framework), QM describes mathematically a state in terms of a superposition (of states). Since the evolution described by QM allows us to predict that the quantum system will get entangled with the apparatus and thus its pointer positions will also become a superposition, the question is why do we observe a single outcome instead of a superposition of them? It is interesting to notice that for Bohr, the measurement problem was never considered. The reason is that through his presuppositions, Bohr begun the analysis of QM presupposing ``right from the start'' classical single outcomes. 
\eq
In fact, what allows us to predict the probabilities of the possible outcomes of a measurement is not the evolution described by QM, for QM describes no evolution. There are at least nine different formulations of QM. The better known among them are Heisenberg's matrix formulation, Schr\"odinger's wave-function formulation, Feynman's path-integral formulation, the density-matrix formulation, and Wigner's phase-space formulation \cite{Styeretal}. Not all of them feature an evolving quantum state. Yet it stands to reason that the interpretation of QM ought to depend on what is common to all formulations of the theory (and thus has a chance of being objective) rather than on the idiosyncrasies of a particular formulation such as Schr\"odinger's.%
\footnote{What is to blame here is the manner in which quantum mechanics is generally taught. While junior-level classical mechanics courses devote a considerable amount of time to different formulations of classical mechanics (such as Newtonian, Lagrangian, Hamiltonian, least action), even graduate-level quantum mechanics courses emphasize the wave-function formulation almost to the exclusion of all variants. This is not only how $\psi$-ontologists come to think of quantum states as evolving physical states but also how QBists come to think of them as evolving states of belief.}

Another reason QM describes no evolution is that the quantum calculus of correlations is time-symmetric. It allows us to assign probabilities not only to the possible outcomes of a later measurement on the basis of an earlier measurement but also to the possible outcomes of an earlier measurement on the basis of a later one.%
\footnote{It even allows us to assign probabilities to the possible outcomes of a measurement on the basis of both earlier and later outcomes using the ABL rule \cite{ABL} rather than Born's.}
It is therefore just as possible to postulate that quantum states evolve backward in time as it is to postulate that they evolve forward in time. If the former postulate contributes nothing to our understanding of QM, then neither does the latter. 

Nor does QM allow us to predict that the quantum system will get entangled with the apparatus pointer---and it had better not, for in the face of overwhelming evidence that measurements tend to have outcomes, this would be absurd. What is common to all formulations of QM is that it serves as a calculus of correlations between measurement outcomes. The reason we observe a single outcome is therefore simply that without single outcomes the quantum calculus of correlations would have no application. There would be nothing to correlate.  ``[P]resupposing `right from the start' classical single outcomes'' is therefore the only sound way to proceed. Thus, contrary to what was claimed by de Ronde, the two (pseudo-) problems fail to make explicit how (or even that) ``QM has turned into a `quantum omelette' with no clear limit between an ontological account and an epistemological one.''

Should (or can) there be a clear limit between the two accounts? Is it even possible to give an ontological account free of any trace of epistemology, or an epistemological account free of any trace of ontology? While it is obviously beyond the scope of the present paper to enter into a discussion of philosophical issues about which countless volumes have been written, off the cuff I would say that no ontological account is complete (or even meaningful) without an epistemological justification, and that no epistemological account is complete (or serves any purpose) if it does not relate to an ontological account (epistemology being about knowledge, and knowledge being about a reality of some kind).

With regard to QM there are two ways to deny this: that of the quantum-state realist, who is cavalier about epistemological concerns, and that of the QBist, who is cavalier about ontological concerns. Nothing much needs to be said about quantum-state realism, inasmuch as this is essentially self-defeating. Any interpretation of QM that needs to account for the existence of measurement outcomes---and thus for the existence of measurements, since no measurement is a measurement if it doesn't have an outcome---is thwarted by the non-objectification theorems proved by Mittelstaedt \cite[Sect.~4.3(b)]{Mittelstaedt} and the insolubility theorem for the objectification problem due to Busch et al.~\cite[Sect.~III.6.2]{BLM}. 

What about QBism? QBists are right in being cavalier about ontological concerns if this means being unconcerned about the relation (if any) between QM and transcendental reality, but they are wrong in being cavalier about the relation between QM and an objective reality. Most if not all of their arguments presuppose such a reality,%
\footnote{For examples the reader is invited to consult my \cite{Mohrhoff_QBism,Mohrhoff_QMExp}.}
whose existence they cannot therefore consistently deny, just as the philosophical skeptic cannot deny a version of realism whose truth she presupposes in defending her stance. To bring home this crucial point, let us assume with Searle \cite[pp.~286--87]{Searle}
\bq
that there is an intelligible discourse shared publicly by different speakers / hearers. We assume that people actually communicate with each other in a public language about public objects and states of affairs in the world. We then show that a condition of the possibility of such communication is some form of direct realism.
\eq
The argument Searle is about to present is directed against the sense-datum theory of perception, according to which all we ever perceive directly---without the mediation of inferential processes---is our own subjective experiences, called ``ideas'' by Locke, ``impressions'' by Hume, and ``representations'' by Kant. In one form or another the sense-datum theory was held by most of the great philosophers in the history of the subject. (QBism may be seen as a throwback to these bygone days.) The argument begins by assuming that we successfully communicate with other human beings at least some of the time, using publicly available meanings in a public language. 
\bq
But in order to succeed in communicating in a public language, we have to assume common, publicly available objects of reference. So, for example, when I use the expression ``this table'' I have to assume that you understand the expression in the same way that I intend it. I have to assume we are both referring to the same table, and when you understand me in my utterance of ``this table'' you take it as referring to the same object you refer to in this context in your utterance of ``this table.''
\eq
The implication is that ``you and I share a perceptual access to one and the same object.'' However, saying that ``you and I are both perceiving the same public object'' does \emph{not} mean that you and I perceive the transcendental object or ``thing in itself.'' The ``direct realism'' Searle is defending is two removes from this na\"\i ve view. By the sense-datum theory we get away from it, but then we realize that 
\bq
Once you claim that we do not see publicly available objects but only sense data, then it looks like solipsism is going to follow rather swiftly. If I can only talk meaningfully about objects that are in principle epistemically available to me, and the only epistemically available objects are private sense data, then there is no way that I can succeed in communicating in a public language, because there is no way that I can share the same object of reference with other speakers.
\eq 
What else is this public language than the ordinary language the necessity of whose use Bohr was at such pains to stress? And what else is the general object of reference of this language than the objective reality which the QBists fail to recognize as the proper object of scientific inquiry, and which de Ronde fails to recognize as the sole reality accessible to scientific inquiry? By throwing out the baby of objective reality with the bathwater of transcendental reality,  QBists have landed themselves on the horns of a dilemma: insofar as they claim to be exclusively concerned with the subjective experiences of individual ``agents'' or ``users,'' they have no way of communicating their views,%
\footnote{This, and not merely the obtuseness of their detractors, appears to be the reason they seem to have such a hard time making themselves understood.}
and insofar as they succeed in communicating, they implicitly acknowledge an objective reality. It no doubt is an interesting project to find out how far the Bayesian interpretation of probability can be carried in the context of QM, but to deny that QM refers to measurement outcomes indicated by instruments situated in an objective reality is overkill. It is an overreaction against the realism of the $\psi$-ontologist, grounded in a common failure to distinguish between the two kinds of reality. 

A result of this failure is the frequent occurrence of the fallacy known as ``false dilemma'': Either we take a transcendental realist stance or we must accept that QM does not make reference to anything but beliefs of ``users.'' Either we embrace $\psi$-ontology or  we ``remain on the surface of intersubjectivity,'' using an epistemic approach that restricts our discourse to the way we interact by communicating empirical findings, leaving aside ``the relation of these interactions to the world and reality themselves'' \cite[p.~7]{CdR}. The possibility which remains unconsidered is that QM makes reference to an objective reality that, while not being the reality of the $\psi$-ontologist, is essential to the expression of our beliefs and the communication of our empirical findings.

De Ronde's ambivalent assessment of QBism reflects the QBists' dilemma. Addressing the horn of solipsism, he writes that QBists
\bq
dissolve all important and interesting questions that physical thought has produced since the origin itself of the theory of quanta. Taking to its most extreme limit several of the main Bohrian ideas, QBism has turned physics into a solipsistic realm of personal experience in which no falsification can be produced; and even more worrying, where there are no physical problems or debates left. QBism does not solve the problems of QM, it simply dissolves them.
\eq
Addressing the other horn (i.e., accepting QBists' ability to communicate in public language), he claims that ``QBists have produced a consistent scheme that might allow us to begin to unscramble---at least part of---the `quantum omelette','' though he gives no indication how this might be done. Nor does he bother to substantiate his extravagant claim that ``QBism has seen much better than Bohr himself the difficult problems involved when applying an epistemological stance to understand QM''---a claim strangely at odds with his statement that ``[e]ven today [Bohr's scheme] seems to us one of the strongest approaches to QM.''

If it were true that ``QBism cannot be proven to be wrong,'' as de Ronde claims, QBism would be \emph{not even wrong}. QBism, however, makes numerous claims, and some of the fourteen examined in my \cite{Mohrhoff_QBism} \emph{are} wrong, for example the claim that there are no external criteria---external to the individual ``user's'' private theater of subjective experiences---for declaring a probability judgment right or wrong. In fact, there are objective data---external to the individual ``user'' though not, of course, external in the transcendental sense---on the basis of which probabilities are assigned, notwithstanding that the choice of these data and hence the probability assignments depend on the ``user.'' 

De Ronde claims to ``show why the epistemic QBist approach is safe from several (ontological) criticisms it has recently received,'' including my own. What appears to have escaped his notice is that none of my criticisms were ontological in his sense. While Marchildon, Nauenberg, and myself are collectively indicted for asking ``QBists to answer ontological questions they have explicitly left aside right from the start,'' he offers not a shred of evidence that his indictment has merit in my case. Our attacks are said to ``come either from the reintroduction of ontological problems,'' which is not true in my case, ``or from the unwillingness to understand the radicalness of the QBist proposal.'' Concerning the latter, I beg forgiveness for quoting from an email I received from Chris Fuchs after posting my \cite{Mohrhoff_QBism}: ``Thanks for your paper tonight. I will read it very carefully in the coming days. Your Section 4 [titled ``The central affirmations of QBism''] so impressed me that I know I *must* read it.'' In a message to his QBist colleagues, forwarded to me in the same email, he further wrote: ``The 14 things he lists in Section 4 are remarkably accurate \dots\ unless I've had too much wine tonight.''

In an attempt at defending $\psi$-ontology, de Ronde points out that ``[t]he foundational discussions that have taken place during the last decades [concerning, among other things, the EPR paper, Bell inequalities, and the Kochen-Specker theorem] are in strict relation to a realist account of the theory.'' If so, what is the conclusion to be drawn from these discussions if not that Bohr was right: realist accounts of QM do not work. In the quantum domain, no property or value is a possessed property or value unless its existence is implied by---indicated by, can be inferred from---an event or state of affairs in the classical domain. The distinction between a classical and a quantum domain is thus an inevitable feature of QM. It needs to be understood, not swept under the rug or explained away.

Again, according to de Ronde, ``[a]ll interesting problems which we have been discussing in the philosophy of science and foundations community for more than a Century \dots\ have been in fact the conditions of possibility for the development of a new quantum technological era''; these problems ``allowed us to produce outstanding developments such as quantum teleportation, quantum cryptography and quantum computation.'' Here de Ronde seems to be speaking off the top of his head, considering that important contributions to these fields came from QBists and other physicists with no transcendental realist leanings.

\section{}
Echoing Kant's famous dictum that ``[t]houghts without content are empty, intuitions without concepts are blind'' \cite[p.~193]{KantCPR}, Bohr could have said that without measurements the formal apparatus of quantum mechanics is empty, while measurements without the formal apparatus of quantum mechanics are blind. What allows us to peer beyond the classical domain with its apparatuses is the combination of measurement outcomes and their quantum-mechanical correlations. And what we find if we peer into the quantum domain is that intrinsically the things we call ``particles'' are identical with each other in the strong sense of numerical identity \cite{Mohrhoff_manifestingQW,Mohrhoff_QMnewlight}. They are one and the same intrinsically undifferentiated Being, transcendent of spatial and temporal distinctions, which by entering into reflexive spatial relations gives rise to
\ben
\item what looks like a multiplicity of relata if the reflexive quality of the relations is ignored, and 
\item what looks like a substantial expanse if the spatial quality of the relations is reified.
\een
In the words of Leibniz: \emph{omnibus ex nihilo ducendis sufficit unum}---one is enough to create everything from nothing. 

As said, the distinction between a classical and a quantum domain needs to be understood, and this (if possible) beyond the linguistic necessity of speaking about the quantum domain in terms of correlations between events in the classical domain. One reason it is so hard to beat sense into QM is that it answers a question we are not in the habit of asking. Instead of asking what the ultimate constituents of matter are and how they interact and combine, we need to broaden our repertoire of explanatory concepts and inquire into the manifestation of the familiar world of everyday experience. Since the kinematical properties of microphysical objects---their positions, momenta, energies, etc.---only exist if and when they are indicated by the behavior of macrophysical objects, microphysical objects cannot play the role of constituent parts. They can only play an instrumental role in the manifestation of macrophysical objects. Essentially, therefore, the distinction between the two domains is a distinction between the \emph{manifested world} and its \emph{manifestation}. 

The manifestation of the familiar world of everyday experience consists in a transition from the undifferentiated state of Being to a state that allows itself to be described in the classical language of interacting objects and causally related events. This transition passes through several stages, across which the world's differentiation into distinguishable regions of space and distinguishable objects with definite properties is progressively realized. There is a stage at which Being presents itself as a multitude of formless particles. This stage is probed by high-energy physics and known to us through correlations between the counterfactual clicks of imagined detectors, i.e., in terms of transition probabilities between in-states and out-states. There are stages that mark the emergence of form, albeit as a type of form that cannot yet be visualized. The forms of nucleons, nuclei, and atoms can only be mathematically described, as probability distributions over abstract spaces of increasingly higher dimensions. At energies low enough for atoms to be stable, it becomes possible to conceive of objects with fixed numbers of components, and these we describe in terms of correlations between the possible outcomes of unperformed measurements. The next stage---closest to the manifested world---contains the first objects with forms that can be visualized---the atomic configurations of molecules. But it is only the final stage---the manifested world---that contains the actual detector clicks and the actual measurement outcomes which have made it possible to discover and study the correlations that govern the quantum domain.

One begins to understand why the general theoretical framework of contemporary physics is a probability calculus, and why the probabilities are assigned to measurement outcomes. If quantum mechanics concerns a transition through which the differentiation of the world into distinguishable objects and distinguishable regions of space is gradually realized, the question arises as to how the intermediate stages are to be described---the stages at which the differentiation is incomplete and the distinguishability between objects or regions of space is only partially realized. The answer is that whatever is not completely distinguishable can only be described by assigning probabilities to what is completely distinguishable, namely to the possible outcomes of a measurement. What is instrumental in the manifestation of the world can only be described in terms of (correlations between) events that happen or could happen in the manifested world.

The atemporal causality by which Being manifests the world must be distinguished from its more familiar temporal cousin. The usefulness of the latter, which links states or events across time or spacetime, is confined to the world drama; it plays no part in setting the stage for it. It helps us make sense of the manifested world as well as of the cognate world of classical physics, but it throws no light on the process of manifestation nor on the quantum correlations that are instrumental in the process. That other causality, on the other hand, throws new light on the nonlocality of QM, which the QBists so nonchalantly dismiss. The atemporal process by which Being enters into reflexive relations and matter and space come into being, is the nonlocal event \emph{par excellence}. Depending on one's point of view, it is either coextensive with spacetime (i.e., completely delocalized) or ``outside'' of spacetime (i.e., not localized at all). Occurring in an anterior relation to space and time, it is the common cause of all correlations, not only of the seemingly inexplicable ones between simultaneous events in different locations but also of the seemingly explicable ones between successive events in the same location.%
\footnote{The diachronic correlations between events in timelike relation are as spooky as the synchronic correlations between events in spacelike relation. While we know how to calculate either kind, we know as little of a physical process by which an event here and now contributes to determine the probability of a \emph{later} event \emph{here} as we know of a physical process by which an event here and now contributes to determine the probability of a \emph{distant} event \emph{now}.}

The objection may be raised that in positing an intrinsically undifferentiated Being and an atemporal process of manifestation, I have ventured into transcendental territory.  But this is not the case.  While Bohr went beyond Kant only in that he opened up the Kantian world-as-we-know-it, providing a window on what lies beyond,%
\footnote{What is responsible for the closure of the objective Kantian world is (i)~Kant's apriorism, which requires (among other things) the universal validity of the law of causality, and (ii)~Kant's principle of thoroughgoing determination, which asserts that ``among all possible predicates of things, insofar as they are compared with their opposites, one must apply to [each thing] as to its possibility'' \cite[p.~553]{KantCPR}.}
I go beyond Bohr only in that I use QM to look through this window. It is still essentially the Kantian categories that I use when speaking of the manifested world as a system of interacting and causally evolving bundles of possessed properties, and it is still the quantum-mechanical correlations between outcome-indicating events in this world that I use to draw my inferences. It is no doubt tempting to think of Being and the manifested world transcendentally, as if they existed out of relation to our experience, but of what exists out of relation to our experience we know zilch. The manifested world exists in relation to our experience---it is manifested \emph{to us}---and so does the Being which manifests it.


\begin{thebibliography}{99}
\bibitem{Jaynes}
E.T. Jaynes: Probability in quantum theory, in \emph{Complexity, Entropy and the Physics of Information}, edited by W.H. Zurek (Addison-Wesley, Redwood City, CA, 1990), pp.\ 381--400.

\bibitem{CdR}
C. de Ronde: QBism, FAPP and the quantum omelette, preprint: arXiv:1608.00548v1 [quant-ph].

\bibitem{Mohrhoff_QBism} 
U. Mohrhoff: QBism: a critical appraisal, preprint: arXiv:1409.3312v1 [quant-ph].

\bibitem{Nauenberg}
M. Nauenberg: Comment on QBism and locality in quantum mechanics, \emph{American Journal of Physics} 83, 197--198 (2015), preprint: arXiv:1502.00123v1 [quant-ph].

\bibitem{FMS_Nauenberg}
C.A. Fuchs, N.D. Mermin, and R. Schack: Reading QBism: a reply to Nauenberg, \emph{American Journal of Physics} 83, 198 (2015).

\bibitem{Bohr-EAPHK}
N. Bohr: \emph{Essays 1958--1962 on Atomic Physics and Human Knowledge} (John Wiley \& Sons, New York, 1963).

\bibitem{Bohr-APHK}
N. Bohr: \emph{Atomic Physics and Human Knowledge} (John Wiley \& Sons, New York, 1958).

\bibitem{UlfBohr}
O. Ulfbeck and A. Bohr: Genuine fortuitousness: where did that click come from? \emph{Foundations of Physics} 31, 757--774 (2001). 

\bibitem{Mohrhoff_clicks}
U. Mohrhoff: Making sense of a world of clicks, \emph{Foundations of Physics} 32, 1295--1311 (2002). 

\bibitem{Bohr1935a}
N. Bohr:  Quantum mechanics and physical reality, \emph{Nature} 136, 65 (1935).

\bibitem{Wheeler}
J.A. Wheeler: Law without law, in \emph{Quantum Theory and Measurement}, edited by J.A. Wheeler and W.H. Zurek (Princeton University Press, Princeton, NJ, 1983) 182--213.

\bibitem{Mohrhoff_manifestingQW}
U. Mohrhoff: Manifesting the quantum world, \emph{Foundations of Physics} 44, 641--677 (2014).

\bibitem{Mohrhoff_QMnewlight}
U. Mohrhoff: Quantum mechanics in a new light. \emph{Foundations of Science} DOI 10.1007/s10699-016-9487-6 (2016); preprint: http://bit.ly/2b5OVY5

\bibitem{Bohr_1939}
N. Bohr: in \emph{New Theories in Physics: Conference Organized in Collaboration with the International Union of Physics and the Polish Intellectual Co-operation Committee}, Warsaw, 30 May--3 June, 1938 (International Institute of Intellectual Co-operation, Paris, 1939), 11--45.

\bibitem{Styeretal}
D.F. Styer, M.S. Balkin, K.M. Becker, M.R. Burns, C.E. Dudley, S.T. Forth, J.S. Gaumer, M.A. Kramer, D.C. Oertel, L.H. Park, M.T. Rinkoski, C.T. Smith, T.D. Wotherspoon: Nine formulations of quantum mechanics, \emph{American Journal of Physics} 70, 288--297 (2002).

\bibitem{ABL} 
Y. Aharonov, P.G. Bergmann, and J.L. Lebowitz: Time symmetry in the quantum process of measurement, \emph{Physical Review B} 134, 1410--1416 (1964).

\bibitem{FMS}
C.A. Fuchs, N.D. Mermin, and R. Schack: An introduction to QBism with an application to the locality of quantum mechanics, \emph{American Journal of Physics} 82, 749--754 (2014).

\bibitem{Mittelstaedt}
P. Mittelstaedt: \emph{The Interpretation of Quantum Mechanics and the
Measurement Process} (Cambridge University Press, Cambridge, MA, 1998).

\bibitem{BLM}
P. Busch, P.J. Lahti, and P. Mittelstaedt: \emph{The Quantum Theory of Measurement}, 2nd Revised Edition (Springer, Berlin, 1996).

\bibitem{Mohrhoff_QMExp}
U. Mohrhoff: Quantum mechanics and experience, preprint: arXiv:1410.5916v2 [quant-ph].

\bibitem{Searle}
J.R. Searle: \emph{Mind: A Brief Introduction} (Oxford University Press, Oxford, UK, 2004).

\bibitem{KantCPR}
I. Kant: \emph{Critique of Pure Reason}, translated and edited by P. Guyer and A.W. Wood (Cambridge University Press, New York, 1999).

\end{thebibliography}
\end{document}